\documentstyle[manuscript,aps,prc,amssymb,tighten,graphicx]{revtex}
\def\be{\begin{equation}}
\def\ee{\end{equation}}
\def\bea{\begin{eqnarray}}
\def\eea{\end{eqnarray}}
\def\bsa{\begin{mathletters}\begin{eqnarray}}
\def\esa{\end{eqnarray}\end{mathletters}}
\def\eps{\varepsilon}
\def\deps{\delta\eps}
\def\ra{\rightarrow}
\def\de{\Delta}
\def\al{\alpha}

\def\om{\omega}
\def\onu{\omega_\nu}
\def\fdag{F^\dagger}
\def\kv{{\bbox{k}}}

\def\lla{\big\langle}
\def\rra{\big\rangle}

\def\drho{\delta\rho}
\def\dmu{\delta\mu}

\def\epk{E^+_\kv}
\def\emk{E^-_\kv}
\begin{document}
\draft

\title{Transition from BCS pairing to Bose-Einstein condensation
       in low-density asymmetric nuclear matter}

%\date{Preliminary draft: \today}

\author{U. Lombardo$^1$, P. Nozi\`{e}res$^2$, P. Schuck$^3$, 
        H.-J. Schulze$^4$, and A. Sedrakian$^3$}
\address{$^1$ Dipartimento di Fisica, 57 Corso Italia, and
         INFN-LNS, Via Santa Sofia, 9500 Catania, Italy}
\address{$^2$ Institut Laue Langevin, BP 156, 38042 Grenoble Cedex 9, France}
%\address{$^2$ CNRS-IN2P3 Universite Joseph Fourier, Institut des Sciences 
%         Nucleaires,\\ 53 Avenue des Martyrs, 38026 Grenoble Cedex, France}
\address{$^3$ Groupe de Physique Theorique, Institut de Physique Nucleaire,
         91406 Orsay Cedex, France}
\address{$^4$ Departament ECM, Universitat de Barcelona,
         Av. Diagonal 647, 08028 Barcelona, Spain}

\maketitle
\begin{abstract}
%Isospin-singlet neutron-proton superfluidity in neutron-rich 
%nuclear matter is studied within the BCS formalism. 
%The excess neutrons are placed in a window around the Fermi surface.
%Therefore a small isospin asymmetry is able to rapidly suppress the 
%pairing gap in the density domain above the Mott threshold. 
%On the contrary, at low density the Bose-Einstein condensation of deuterons is
%little affected by an additional gas of free neutrons even at 
%large asymmetries.
We study the isospin-singlet neutron-proton pairing in bulk nuclear matter 
as a function of density and isospin asymmetry within the BCS formalism. 
In the high-density, weak-coupling regime the neutron-proton paired state 
is strongly suppressed by a minor neutron excess. 
As the system is diluted, the BCS state with large, overlapping Cooper 
pairs evolves smoothly into a Bose-Einstein condensate of tightly bound 
neutron-proton pairs (deuterons). 
%The crossover occurs in the vicinity of the Mott transition point, 
%where the combined chemical potential of the constituents becomes negative. 
In the resulting low-density system a neutron excess is ineffective in 
quenching the pair correlations because of the large spatial separation 
of the deuterons and neutrons.
As a result, the Bose-Einstein condensation of deuterons is weakly affected 
by an additional gas of free neutrons even at very large asymmetries. 
\end{abstract}


%------------------------------------------------------------------------------
\section{Introduction}
\label{s:intro}

The crossover from BCS superconductivity to Bose-Einstein
condensation (BEC) manifests itself in fermionic systems with attractive 
interactions whenever either the density is decreased and/or
the interaction strength in the system is increased sufficiently. 
The transition from large overlapping Cooper pairs to tightly bound 
non-overlapping bosons can be described entirely within the ordinary BCS 
theory, if the effects of fluctuations are ignored (mean-field approximation). 
%Surprisingly, the BEC limit at the mean-field level is well decribed by the 
%BCS theory; 
Indeed in the free-space limit the gap equation reduces 
to the Schr\"odinger equation for bound pairs. 
%i.e., Cooper's theorem remains intact even when the notion of the Fermi 
%surface loses its meaning as the system is diluted.
This type of transition has been studied initially in the context of ordinary
superconductors \cite{LEG}, excitonic superconductivity in 
semiconductors \cite{KEL}, and, at finite temperature, in an attractive 
fermion gas \cite{NOZ}. 
Although the BCS and BEC limits are physically quite different, 
the transition between them was found smooth within the ordinary BCS theory.

More recently it was argued \cite{ALM,SSALR,SCK} 
that a similar situation should occur in symmetric nuclear matter, where
neutron-proton ($np$) pairing undergoes a smooth transition from a state of
$np$ Cooper pairs at higher densities to a gas of Bose-condensed deuterons, 
when the nucleon density is reduced to extremely low values. 
At the same time the chemical potential evolves from positive values to 
negative ones (disregarding a mean field), approaching half of the deuteron 
binding energy in the zero-density limit. 
This transition may be relevant, and could give valuable information on 
$np$ correlations, in low-density nuclear systems like the surface of nuclei,
expanding nuclear matter from heavy ion collisions, collapsing stars, etc.

The $np$ pairing effect is largest in the isospin symmetric systems. 
However, most of the systems of interest are, to some extent, 
isospin asymmetric, i.e., there is usually a certain excess of neutrons
over protons. 
In this case the pairing is suppressed, since for neutrons and protons 
lying on different Fermi surfaces the phase space overlap decreases as 
these surfaces are pushed further apart by isospin asymmetry 
\cite{ALM,ARM1,ARM2}.
Note that this situation is quite analogous to pairing of, e.g., 
spin-polarized electrons. 
As is well known, an extra amount of spin-up electrons over the spin-down 
electrons is very efficient in destroying the superfluidity in the 
weak-coupling regime.

This situation is well known also from odd nuclei or nuclei with 
quasi-particle excitations, where the so-called blocked 
gap equation emerges \cite{RS}. 
Indeed, the ground state wave function for a $np$ BCS state with $N$ excess 
neutrons can be written in the following way \cite{RS},
\bea
 \big| \Phi \rra_N &=& 
 \al^\dag_{n\kv_1} \al^\dag_{n\kv_2} \ldots \al^\dag_{n\kv_N} 
 \big|{\rm BCS}\rra
\nonumber\\
 &=&
% a^\dag_{n\kv_1} a^\dag_{n\kv_2} \ldots a^\dag_{n\kv_N} 
 \prod_{\kv = \kv_1, \ldots, \kv_N}\!\! a^\dag_{n\kv} 
 \prod_{\kv \ne \kv_1, \ldots, \kv_N}\!\!
 \left( u_{\kv} + v_{\kv} a^\dag_{n\kv} a^\dag_{p-\kv} \right) 
 \big|{\rm vac}\rra 
 \:, 
\label{e:gs}
\eea
where $\alpha^\dag$ and $a^\dag$ are the quasiparticle and bare particle
creation operators, and $\big| {\rm BCS} \rra$ is the superfluid ground state
in the symmetric case. 
From Eq.~(\ref{e:gs}) we see that the presence of the extra neutrons entails a
suppression of the Cooper pairs in the $np$ BCS state, which have the same 
quantum numbers as the extra neutrons. 
The variational minimization of the Hamiltonian then leads to the so-called 
blocked gap equation \cite{RS}, where the window of $\kv_1, \ldots, \kv_N$ 
states is missing in the integral equation for the gap. 
Since this window is situated close to the Fermi energy,
where most of the pairing correlations are built, 
the suppression mechanism is extremely efficient.
This is the case for positive chemical potential. 

The situation is, however, more complex when the chemical potential 
approaches zero and becomes negative, i.e., in the limit of strong coupling. 
Clearly the $np$ pairs will evolve to tightly bound deuterons in this limit,
and, because of the isospin asymmetry, they will coexist with
a dilute gas of excess neutrons.  
It is physically conceivable that in this limit,
where the deuterons as well as the extra neutrons are extremely dilute, 
the Pauli blocking effect of neutrons on the deuterons will become negligeable.
This may have interesting consequences, for example, in the far tail of 
nuclei, where a deuteron condensate may exist in spite of the fact that 
there the density can be quite asymmetric. 
Apart from finite nuclei, similar physical effects could play an important 
role in other low-density asymmetric nuclear systems. 

The aim of this work is, therefore, a detailed study of the behavior of $np$ 
pairing in asymmetric nuclear matter as a function of density and asymmetry. 
The paper is organized as follows. 
In section~\ref{s:form} we set up the general frame in terms of
the Gorkov formalism at finite temperature.
In section~\ref{s:zero} we specify the relevant equations to zero temperature 
and discuss the analytical limiting case in the weak-coupling approximation.
Numerical results are presented and discussed in section~\ref{s:num}.
Our conclusions are summarized in section~\ref{s:conc}.

%------------------------------------------------------------------------------
\section{Basic equations}
\label{s:form}

%\subsection{Green's function formalism for an isospin-asymmetric superfluid}

In this section we briefly review the treatment of an isospin-singlet 
superfluid Fermi system within the Green's function formalism, 
see also Refs.~\cite{SCK,ARM1,ARM2}.  
We use the standard method of Green's functions at finite temperatures,
see, e.g., Refs.~\cite{ABRI,NOZ63,NOZ66,SCHRIEF,MIG}.
In this formalism interacting superfluid systems are described in terms of 
the Gorkov equations, which generalize the Dyson equation for normal Fermi
systems by doubling the number of propagators. 
For a homogeneous system the normal and anomalous propagators, $G$ and $F$, 
are defined as
%The principal equations describing a superfluid system are the Gorkov
%equations, which can be considered a generalization of the Dyson equation 
%for a normal Fermi system.
%They express the relation between normal and anomalous propagators 
%$G$ and $F$, defined for a homogeneous system as
\bsa
 G_{\tau\sigma,\tau'\sigma'}(\kv,t-t') &=&
 {1\over i} \lla T \big[ 
 \psi_{\tau\sigma}(\kv,t) \psi_{\tau'\sigma'}^\dag(\kv,t') \big] \rra  \:,
\\
 F_{\tau\sigma,\tau'\sigma'}(\kv,t-t') &=&
 {1\over i} \lla T \big[
 \psi_{\tau\sigma}(\kv,t) \psi_{\tau'\sigma'}(\kv,t') \big] \rra  \:, 
\esa
where $\sigma =\;\uparrow,\downarrow$ and $\tau=n,p$ 
denote the spin and isospin quantum numbers, respectively.
On introducing the Matsubara frequencies $\onu = (2\nu+1)\pi T$,
where $T\equiv 1/\beta$ is the temperature, $\nu = 0, 1, 2, \dots $,
%Introducing, for the treatment at finite temperature $T=1/\beta$,
%the Matsubara frequencies $\onu = (2\nu+1)\pi T$, 
the propagators are conveniently written in the Fourier representation as
\bsa
 G_{\tau\sigma,\tau'\sigma'}(\kv,t) &=& {1\over\beta} 
 \sum_\nu e^{-i\onu t} G_{\tau\sigma,\tau'\sigma'}(\kv,\onu) 
 %\delta_{\tau\tau'}  
 \:,
\\
 F_{\tau\sigma,\tau'\sigma'}(\kv,t) &=& {1\over\beta} 
 \sum_\nu e^{-i\onu t} F_{\tau\sigma,\tau'\sigma'}(\kv,\onu) 
 %(1-\delta_{\tau\tau'})  
 \:.
\label{s:green}
\esa
Using the short-hand notation $\bf D, \de, G$, and $\bf F$, for $4\times4$ 
matrices in spin-isospin space, the Gorkov equations can be written as  
%They fulfill equations of motion written in terms of $4 \times 4$ spin-isospin
%matrices $\bf D, \de, G, F$,
\be
 \left( \begin{array}{cc}
  {\bf D}(\kv,\onu)  & {\bf\de}_{}(\kv)  \\ 
  {\bf\de}_{}^\dagger(\kv) & -{\bf D}(\kv,-\onu)
 \end{array} \right)
 \left( \begin{array}{c} 
  {\bf G}_{}(\kv,\onu) \\ {\bf F}^\dag_{}(\kv,\onu)
 \end{array} \right) =
 \left( \begin{array}{c} 
  {\bf 1} \\ {\bf 0} 
 \end{array} \right) \:, 
\label{e:fg}
\ee
where
\be
  {\bf D}(\kv,\onu) = 
 \left( \begin{array}{cccc}
  i\onu - \eps_{n\uparrow}(\kv) & 0 & 0 & 0 \\ 
  0 & i\onu - \eps_{n\downarrow}(\kv) & 0 & 0 \\
  0 & 0 & i\onu - \eps_{p\uparrow}(\kv) & 0 \\
  0 & 0 & 0 & i\onu - \eps_{p\downarrow}(\kv) 
 \end{array} \right)
\ee
and $\bf 1$ denotes the four-dimensional unit matrix.
Since we are mostly interested in the pairing properties at very low density,
where the nuclear mean field plays a minor role, 
the single-particle spectrum adopted in this work is simply the kinetic one,
\bsa
 \eps_{n\uparrow}(\kv) = \eps_{n\downarrow}(\kv) \equiv \eps_n(\kv) &=& 
 {\kv^2 \over 2m} - \mu_n = \eps_\kv + \delta\mu  \:,
\\
 \eps_{p\uparrow}(\kv) = \eps_{p\downarrow}(\kv) \equiv \eps_p(\kv) &=& 
 {\kv^2 \over 2m} - \mu_p = \eps_\kv - \delta\mu  \:,
\esa   
where $\eps_\kv={\kv^2\! / 2m} -\mu$, $\mu = (\mu_n+\mu_p)/2$ is the 
average chemical potential between protons and neutrons, 
and $\delta\mu = (\mu_n-\mu_p)/2$ is the associated shift. 
From now on we assume nuclear matter with neutron excess 
so that $\delta\mu > 0$.             

In this article we concentrate on pairing in the dominant isospin-singlet
channel
%spin-triplet channel $^3SD_1$, 
that can be described by a gap function ${\bf\de}(\kv)$ with the structure 
\be
  {\bf\de} = 
 \left( \begin{array}{cccc}
  0 & 0 & 0 & \de_0+i \de_1 \\ 
  0 & 0 & -\de_0+i \de_1 & 0 \\
  0 & -\de_0-i \de_1 & 0 & 0 \\
  \de_0-i \de_1 & 0 & 0 & 0 
 \end{array} \right)
 \:,
\ee
which is a particular case 
%If the ground state is assumed to be time-reversal invariant, 
%the gap function has the structure 
of a unitary state \cite{SCK,TIL,BALDO},
\be
 {\bf\de}^\dag {\bf\de} = \Delta^2 {\bf 1}  
 \ , \quad
 \Delta^2 = \sqrt{\det{\bf\de}} = \de_0^2 + \de_1^2 
 \:.
\ee
It allows in principle the coexistence of spin-singlet ($S=0$) and 
spin-triplet ($S=1$) pairing correlations $\de_S$.
In the low-density region that we are interested in, pairing is realized
in the spin-triplet $s$-wave channel $^3SD_1$, i.e., $\de = i\de_1$.

\begin{figure}
\includegraphics[totalheight=6.cm,angle=90,bb= 30 500 510 1090]{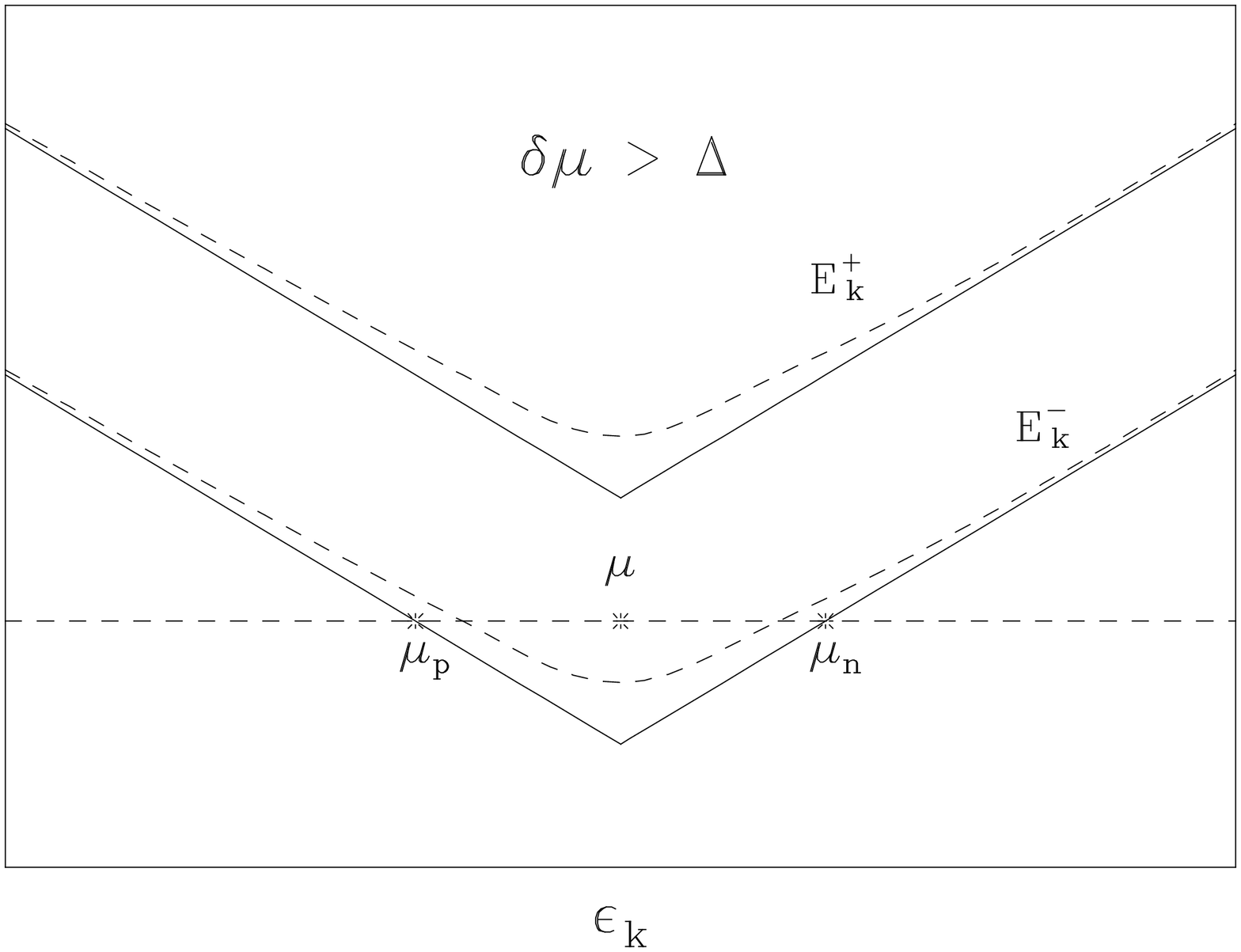}
\caption{
Sketch of the quasiparticle spectrum.
For $\epsilon<\mu$ the lower (upper) solid line correponds to 
neutrons (protons) in the normal state. 
For $\epsilon>\mu$ the lower (upper) solid line corresponds to 
protons (neutrons) in the normal state. 
The corresponding dashed lines refer to nucleons in the superfluid state.
}
\label{f:qp}
\end{figure}

The anomalous propagator ${{\bf\fdag}}$ has the same spin-isospin structure 
as ${\bf\de}$, whereas the normal propagator $\bf G$ is diagonal in the 
spin-isospin indices.
Taking this into account, the system Eq.~(\ref{e:fg}) can be inverted with 
the solution
\begin{mathletters}
\bea
 G_n(\kv,\onu) &=& {i\onu + \eps_p(\kv) \over
 \left( i\onu - \epk \right) \left( i\onu + \emk \right) } \:,
\label{e:G1n}
\\
 G_p(\kv,\onu) &=& {i\onu + \eps_n(\kv) \over
 \left( i\onu - \emk \right) \left( i\onu + \epk \right) } \:,
\label{e:G1p}
\\
 F_{np}^\dag(\kv,\onu) &=& {-\de(\kv) \over
 \left( i\onu - \emk \right) \left( i\onu + \epk \right) } \:,
\label{e:G21}
\\
 F_{pn}^\dag(\kv,\onu) &=& {+\de(\kv) \over
 \left( i\onu - \epk \right) \left( i\onu + \emk \right) } \:,
\label{e:G22}
\eea
\end{mathletters}
where
\be
 E^\pm_\kv = E_\kv \pm \dmu 
 = \sqrt{ \eps_\kv^2 + \Delta(\kv)^2} \pm \dmu \:.
\label{e:psp}
\ee
%the upper sign corresponding to neutrons and the lower to protons. 
The isospin asymmetry thus lifts the degeneracy of the quasiparticle spectra,
leading to two separate branches for protons and neutrons.
This is sketched in Fig.~\ref{f:qp}.
In the region around $\mu$, where $E_\kv < \dmu$, 
the superfluid state is unstable against the normal state
and indeed it accomodates the unpaired neutrons.
This ``isospin window'' is responsible for the pairing gap suppression 
in the density domain where nuclear matter is superfluid. 
It corresponds to the blocking effect for the pairing in nuclei \cite{RS}.   

After analytical continuation of the Green's functions in the complex 
$\om$-plane one calculates the spectral function $A(\kv,\om)$, 
which is given by the discontinuity of $G$ across the real axis, 
\be
 A_\tau(\kv,\om) = 2\pi \left[ 
 \frac{1}{2} \Big( 1 + \frac{\eps_\kv}{E_\kv} \Big) \delta(\om - E^\pm_\kv) +
 \frac{1}{2} \Big( 1 - \frac{\eps_\kv}{E_\kv} \Big) \delta(\om + E^\mp_\kv) 
 \right] \:, 
\ee
and then the density of particles 
\bea
 \rho_\tau &=& 2 \sum_k \int\!\frac{d\om}{(2\pi)} A_\tau(\kv,\om) f(\om) 
\nonumber\\  
 &=& 2 \sum_k \left[
 \frac{1}{2} \Bigl( 1+\frac{\eps_\kv}{E_\kv} \Bigr) f(E^\pm_{\kv}) +
 \frac{1}{2} \Bigl( 1-\frac{\eps_\kv}{E_\kv} \Bigr)
 \left[ 1 - f(E^\mp_{\kv}) \right] \right] \:,
\label{e:dens}
\eea      
where $\sum_k = \int\!d^3k/(2\pi)^3$ and
$f(E) = [1+\exp(\beta E)]^{-1}$ is the Fermi distribution function.
%$\beta^{-1}=k_BT$, $T$ is the temperature, 
%and $k_B$ is the Boltzmann constant.

Analogously, the discontinuity of the anomalous propagator $F^\dag$, 
\be
 B(\kv,\om) = 2\pi 
 \left[ \delta(\om - E^+_\kv) - \delta(\om + E^-_\kv) \right] 
 {\Delta(\kv) \over 2E_\kv } \:,
\ee
yields the gap equation
\bea
 \Delta(\kv') &=& \sum_k \lla\kv'|V|\kv\rra
 \int\!\frac{d\om}{2\pi} B(\kv,\om) f(\om) 
\nonumber\\
 &=& -\sum_k  \lla\kv'|V|\kv\rra 
 {\Delta(\kv) \over 2E_\kv } 
 \left[ 1 - f(\epk) - f(\emk) \right] \:.
\eea
Using the angle-averaging procedure, which is an adequate
approximation for the present purpose (see Ref.~\cite{BALDO}),
the BCS gap equation for asymmetric nuclear matter can be derived:
\be
 \Delta_{l'}(k') = -\sum_{k}\sum_{l}
 V_{ll'}(k,k') {\Delta_{l}(k) \over 2E_k }
 \left[ 1 - f(E^+_k) - f(E^-_k) \right] 
 \ ,\ (l,l'=0,2)
 \:,
\label{e:gap}
\ee
where $E_k^2 = \eps_k^2 + \de(k)^2$
and $\de(k)^2 \equiv \Delta_0(k)^2+\Delta_2(k)^2$ 
is the angle-averaged neutron-proton gap function.
The driving term, $V_{ll'}$, is the bare interaction in the 
$^3SD_1$ channel.
We use for the numerical computations in this work the Argonne $V_{14}$ 
potential \cite{V14}.

From Eq.~(\ref{e:dens}) the total density $\rho =\rho_n+\rho_p$ 
and the neutron excess $\delta\rho =\rho_n-\rho_p$ can easily be derived
and one gets
\begin{mathletters}
\bea
 \rho &=& 2\sum_{k} n_k
 \ ,\quad
 n_k = 1 - \frac{\eps_k}{E_k} \left[ 1 - f(E^+_k) - f(E^-_k) \right] \:,
\label{e:rho}
\\
 \delta\rho &=& 2\sum_k \delta n_k
 \ ,\quad
 \delta n_k = f(E^-_k) - f(E^+_k) \:.
\label{e:drho}
\eea
\label{e:xrho}
\end{mathletters}
For a convenient parametrization of the total density, we also introduce
the quantity
$k_F \equiv (3\pi^2\rho/2)^{1/3}$, 
which is however, 
apart from the isospin symmetric system,
not to be identified with a Fermi momentum.

Introducing the anomalous density 
\be
 \psi_l(k) = \lla a^\dag_{n,\kv} a^\dag_{p,-\kv} \rra_l
 = \frac{\Delta_l(k)}{2E_k}
 \left[ 1 - f(E^+_k) - f(E^-_k) \right] \:,
\label{e:corr}
\ee
and making use of Eq.~(\ref{e:rho}), 
the gap equation, Eq.~(\ref{e:gap}), 
can be recast in the Schr\"odinger-like form
\be
 %\left[ 1 - f(E^+_{k}) - f(E^-_{k}) \right]
 {k^2\over m} \psi_l(k)
 + \left( 1 - n_k \right)
 \sum_{k'}\sum_{l'} V_{ll'}(k,k') \psi_{l'}(k')
 = 2 \mu\, \psi_{l}(k)
 %= -2E_k \psi_{l}(k)
\:.
\label{e:schr}
\ee
In the limit of vanishing density, $n_k\ra0$, 
this equation goes over into the 
Schr\"odinger equation for the deuteron bound state\cite{SCK}.
The chemical potential $2\mu=\mu_p+\mu_n$ 
%contained in $E_k$ 
plays then the role of the energy eigenvalue.

Let us finally remark that in this article we do not study the competition
between and possible coexistence of isospin singlet and triplet 
pairing \cite{AG}. 
It is clear that $T=1$ $nn$ and $pp$ pairing gaps, which are unaffected by 
the isospin asymmetry, will be larger than the $T=0$ gap at sufficiently 
large asymmetry. 
Our results are thus valid up to the asymmetries (yet unspecified) at which 
the isospin triplet pairing becomes dominant. 
On the other hand, as we show below, for very low density systems the 
suppression mechanism due to the shift of Fermi surfaces is ineffective 
and it is reasonable to assume that the $np$ pairing will dominate other 
channels in a wide range of values of the asymmetry.
%However, the present article is dedicated to the model study of pure $T=0$
%pairing, in particular at very low density, where this is indeed the 
%dominant channel.

We also restrict ourselves to work with a free single-particle spectrum, 
which is adequate at low density.
At higher density, when the effective mass deviates substantially from
the bare mass, renormalization of the single-particle spectrum 
should be taken into account \cite{BALDO,BHF,DIS}. 
%a more realistic single-particle spectrum should be used \cite{BALDO,BHF},
%and also dispersive effects \cite{DIS} are expected to become important.
This affects the magnitude of the pairing gap in general and the critical 
asymmetries at which the pairing effect disappears \cite{ARM2}.
In view of this approximation, the results shown below should be 
considered as only qualitative in the high-density region.

%polarization ???

%------------------------------------------------------------------------------
\section{Zero temperature gap equation}
\label{s:zero}

In order to disentangle the isospin effects from the thermal ones, 
we focus in the following on the limit of zero temperature, 
where $f(E^+_k)=0$ and
$f(E^-_k) = 1 - \theta(E^-_k)$, $\theta$ being the step function.
In this limit Eqs.~(\ref{e:gap}) and (\ref{e:xrho}) 
%(\ref{e:rho}), and (\ref{e:drho}) 
reduce to the following ones,
\begin{mathletters}
\bea
 \Delta_{l'}(k') &=& -\sum_{k}\sum_{l}
 V_{ll'}(k,k'){\Delta_{l}(k) \over 2E_k } \theta(E^-_k) \:,
\label{e:gap0} 
\\
 \rho &=& 2\sum_{k} \Bigl[ 1 - \frac{\eps_k}{E_k} \theta(E^-_k) \Bigr] \:,
\label{e:rho0}
\\
 \delta\rho &=& 2\sum_{k} \Bigl[ 1 - \theta(E^-_k) \Bigr] \:.
\label{e:drho0}
\eea
\label{e:eq0} 
\end{mathletters}
In general these three coupled nonlinear equations have to be solved
numerically, maintaining the self-consistency. 
Before presenting the numerical results, let us however first discuss in 
some detail the physical content of these equations.
The physical interpretation of the formalism is that in asymmetric nuclear 
matter a superfluid state of $np$ Cooper pairs with density 
$\rho-\delta\rho = 2\rho_p$ 
coexists with a gas of free neutrons with density $\delta\rho$. 
Since this latter is occupying a region around the Fermi surface 
[determined by $1-\theta(E^-_k)$], 
the momentum space available for the pairing is reduced. 
Thus the effect of isospin asymmetry is to reduce the magnitude of the 
energy gap; 
with increasing neutron excess the superfluidity rapidly 
disappears \cite{ALM,ARM1,ARM2}.

The solution of the gap equation, Eq.~(\ref{e:eq0}), 
assumes different properties according to the value of the chemical potential.
One may distinguish three domains: 
(i) $\mu \gg \de$ (weak-coupling, pairing regime), 
(ii) $\mu \approx 0$ (Mott transition), and 
(iii) $\mu < 0$ (strong-coupling, bound-state regime).
Approximate analytical results can be found in the cases (i) and (iii),
and are described in the following:

\subsubsection*{(i) Weak-coupling regime}

In the region $\mu \gg \de$ the gap equation in the form of Eq.~(\ref{e:gap0})
can be solved in the usual weak-coupling approximation \cite{FW},
in order to gain some insight in the qualitative behavior of the pairing 
properties in the asymmetric case.
This approximation is not very accurate, but it preserves the 
physical content of the exact solution.

\begin{figure}
\includegraphics[totalheight=4.cm,angle=0,bb=-130 520 -130 740]{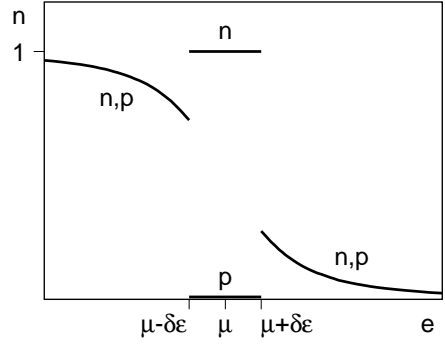}
\caption{
Momentum distributions of neutrons and protons in asymmetric 
superfluid matter.}
\label{f:dis}
\end{figure}

First, one notes from Eq.~(\ref{e:drho0}) that 
the unpaired neutrons are concentrated in the energy interval 
$[\mu-\deps ,\mu+\deps]$, 
with the half-width
\be
 \deps = \sqrt{\dmu^2-\de^2} \:.
\ee
This interval 
is free of protons and does not contribute to the pairing interaction, whereas
outside of it neutron and proton distributions are equal and
given by the BCS result, see Eq.~(\ref{e:rho0}).
This situation is sketched in Fig.~\ref{f:dis}.
One notes at this point that the condition $\dmu>\de$ has to be fulfilled
in order to generate an asymmetry. 
Indeed for $\de,\deps \ll \mu$ the momentum distributions of the two species 
are very sharp 
(practically Fermi distributions)
and one obtains as relation between the asymmetry $\al$ and $\deps$:
\be
 \alpha = {\rho_n-\rho_p \over \rho_n+\rho_p}
 \approx { \left( \int_0^{\mu+\deps}-\int_0^{\mu-\deps} \right) de \sqrt{e}  
 \over \left( \int_0^{\mu+\deps}+\int_0^{\mu-\deps} \right) de \sqrt{e}  } 
 \approx {3\over2}{\deps\over\mu}  
 %\ , \quad (\deps \ll \mu) 
 \:,
\label{e:al}
\ee
i.e., the asymmetry is directly proportional to $\deps$.
Thus we see that $\deps$ has the interpretation of an ``effective''
difference of chemical potentials, since it is $\deps$ and not $\dmu$
which determines the density of excess neutrons for $\de \neq 0$.
Only in the normal system one has $\deps=\dmu$, whereas pairing screens
$\dmu$ such that $\deps<\dmu$, and it needs a critical finite $\dmu=\de$
in order to obtain a finite density of excess neutrons.

\begin{figure}
\includegraphics[totalheight=7.cm,angle=0,bb=-180 290 -180 750]{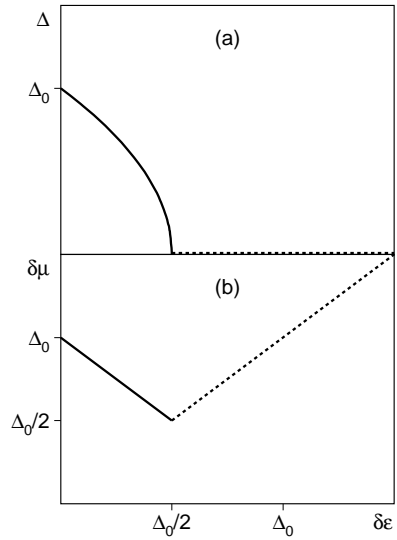}
\caption{
Dependence of gap $\de$ (a) and 
%width of the blocking interval $\deps$ 
%on the chemical potential difference $\dmu$ 
chemical potential difference $\dmu$ (b)
on the width of the blocking interval $\deps$ 
within the weak-coupling 
approximation, see Eq.~(\ref{e:de}). % and (\ref{e:wca}).
The solid lines denote the superfluid phase 
and the dashed lines the normal one.}
\label{f:wc}
\end{figure}

Coming now to the gap equation, 
with a contact interaction $V$ and energy cutoff $\eps_c$, 
Eq.~(\ref{e:gap0}) reads
\be
 %-{8\pi^2 \over (2m)^{3/2}V}  = 
 %\int_0^{e_c} \!\!de\, \sqrt{e\over (e-\mu)^2 + \de^2}
 %\;\theta\Big[ \sqrt{(e-\mu)^2 + \de^2} - \dmu \Big] \:.
 1 = -{(2m)^{3/2}V \over 8\pi^2} 
 \int_{-\mu}^{\eps_c} \!\!d\eps 
 \,\theta\big( |\eps| - \deps \big) 
 \sqrt{ \eps+\mu \over \eps^2 + \de^2}
  %\;\theta\Big[ \sqrt{\eps^2 + \de^2} - \dmu \Big] \:,
 \:.
\label{e:wc}
\ee
For $\de \ll \mu \ll \eps_c$, the integral 
%in Eq.~(\ref{e:wc})
can be approximately calculated, yielding
\be
 1 = N(0)V \ln\left[ { \deps+\dmu \over 2\sqrt{\mu \eps_c} } \right] \:,
\ee
where 
%$N(0)=3\rho/2\mu$ 
$N(0)=m k_F/2\pi^2$ 
is the level density at the Fermi surface of one type of particles.
Therefore
\be
 \dmu + \deps = {\rm const.} = \de_0 
 \quad \Leftrightarrow \quad 
 \de^2 = \de_0^2 - 2\de_0\deps 
% \de^2 = \de_0 \bigl[ 2\delta\mu - \de_0 \bigr] 
% = \de_0 \bigl[ \de_0 - 2\delta\rho/N(0) \bigr]
 \:,
\label{e:de}
\ee
where $\de_0$ is the value of the gap in the symmetric system of the same
density.
We again see that it is $\deps$, and not $\dmu$, which determines 
the reduction of $\de$ from its symmetric value.
The gap decreases very rapidly and disappears when the width of the window
reaches $2\deps = \de_0$, i.e., the size of the gap at symmetry. 
This is shown in Fig.~\ref{f:wc}(a).

We can now combine Eqs.~(\ref{e:al}) and (\ref{e:de})
in order to specify the dependence of the gap on the asymmetry:
\be
 {\de\over\de_0} = \sqrt{ 1 - {4\mu\over3\de_0}\al } \:.
\label{e:wcaa}
\ee
The superfluidity vanishes smoothly, 
but with an infinite slope at
\be
 \al_{\rm max} = {3\de_0 \over 4\mu} \:,
\label{e:amax}
\ee
which in the weak-coupling limit is a very small value.

%Therefore 
We can also determine the dependence of the gap on the 
difference of chemical potentials $\dmu$ \cite{ARM1,STO}:
\be
 {\de\over \de_0} = \sqrt{{2\dmu \over \de_0} - 1} \:.
\label{e:wca}
\ee
%where $\de_0$ is the gap in symmetric matter.
In the symmetric system one has $\dmu=\de_0$, $\deps=0$,
while increasing the asymmetry
the gap decreases with {\em decreasing $\dmu$} and increasing $\deps$ 
and vanishes at $\dmu = \deps = \de_0/2$.

Physically, this behavior can be understood by noting that 
the chemical potential difference $2\dmu=\mu_n-\mu_p$
is the energy one must invest in order to remove a proton from the system 
and then to insert a neutron instead, in other words the symmetry energy.
However, in order to do so in the superfluid system with a neutron excess
(even very small), one must necessarily break a pair in order
to remove the proton, but then one does not gain energy 
by adding the neutron to the top of its Fermi sea.
That is why $\dmu \geq \de$ in the pairing regime  
and $\dmu \geq \sqrt{\mu^2+\de^2}$ in the bound state regime
(the equalities holding in the symmetric system).
So, clearly in a superfluid system of any density and asymmetry,
$\dmu$ can never be zero, because it involves breaking a 
pair (or bound state).

Then, successively increasing the asymmetry by breaking pairs in this way
and filling up the Fermi sea of excess neutrons,
the pairing correlations are destroyed, and consequently one must invest 
less energy in order to exchange $p \leftrightarrow n$.
Therefore $\dmu$ {\em decreases} with increasing asymmetry 
(while $\delta\eps$ increases), up to the point where, 
at $\dmu=\deps=\de_0/2$, 
the $np$ gap disappears completely and 
$2 \dmu$ is simply given by the difference of Fermi energies of the two 
noninteracting Fermi gases at this asymmetry.
The system is then in the normal phase, $\deps=\dmu$, which means that 
in order to further increase the asymmetry the parameter $\dmu$ has to
{\em increase} again, up to $\dmu=\mu$, corresponding to pure neutron
matter.
Consequently, values of $\dmu$ between $\de_0/2$ and $\de_0$ are present 
in the superfluid as well as in the normal phase,
corresponding to different compositions of small and large asymmetry, 
respectively. 
The relations between $\de$, $\dmu$, and $\deps$ are sketched 
in Fig.~\ref{f:wc}. 

We comment finally on the analogy to an electron system in a magnetic field.
Whereas in the case of nuclear matter discussed before, an increasing 
asymmetry is imposed on the system, leading to a smooth disappearance of the
gap at a certain maximum asymmetry, the situation for the electron system
is qualitatively different:
Varying in this case the magnetic field is equivalent to imposing the 
chemical potential difference $\dmu$ instead of the asymmetry,
which now corresponds to the magnetization of the system.
Therefore one observes in this case a first-order phase transition 
at $\dmu=\de_0$, when the 
pairs are suddenly broken up and 
the system jumps immediately from the symmetric superfluid phase
to the magnetized normal phase.
A magnetized superfluid phase is thermodynamically unstable in this case.

\subsubsection*{(ii) Mott-transition regime}

Let us now discuss the situation when the density $\rho$ decreases. 
As we have shown in \cite{SCK}, simultaneously the chemical potential $\mu$
decreases (disregarding the Hartree-Fock shift) 
%The chemical potential decreases (disregarding the Hartree-Fock shift) 
%as a function of density 
and at a certain (very low) density 
(of the order of $\rho \approx \rho_0/100$, 
where $\rho_0 \approx 0.17\;\rm fm^{-3}$ is the nuclear matter 
saturation density) $\mu$ passes through zero, 
corresponding to the Mott transition of the deuteron \cite{SCK}.
This also remains true in the asymmetric case; in fact,
the weak-coupling approximation used above predicts that in pure neutron matter
this happens at $\mu_n=(\pi/2)^2 \mu_D$, 
where $\mu_D=E_D/2 \approx 1.1\;\rm MeV$ is half of the deuteron 
binding energy.
This corresponds to a density $\rho\approx0.0016\;\rm fm^{-3}$.

It seems clear that the Pauli blocking starts to loose its efficiency 
once the left-hand side of the window $[\mu-\deps ,\mu+\deps]$
passes beyond zero, 
because then only part of the window participates in the blocking.
When $\mu$ has reached negative values, less than half of the blocking window
is actually effective. 
While in the symmetric case nothing particular happens when the $np$ Cooper 
pairs change their character to bound deuterons in the low-density limit, 
the asymmetric case 
%with $\delta \mu \ne 0$ 
needs special attention. 
Here both $\delta\mu$ and $\mu$ vary strongly with asymmetry
(at fixed total density),
and the results have to be found numerically, solving the coupled 
Eqs.~(\ref{e:eq0}).
%Eqs.~(\ref{e:gap0}-\ref{e:drho0}).

%The vanishing of $\mu$ marks the Mott transition, i.e., the transition of the 
%nucleon system from the pairing regime to a deuteron Bose condensate. 
%The Mott transition is a smooth one in symmetric nuclear matter \cite{SCK}.
%But in close proximity of $\mu=0$ a quite small asymmetry is enough to rapidly
%suppress the gap, since the typical enhancement of the abnormal density around
%the Fermi surface is removed by an equally narrow isospin window. 
%The gap suddenly jumps from its symmetric value $\de_0$ to zero.
%Imposing a small temperature is able to restore the smooth character of the
%transition (as shown in Fig.~\ref{f:gap}). 

\subsubsection*{(iii) Strong-coupling regime}

Once the density $\rho$ becomes so low 
that simultaneously the mean distance among the deuterons as well as 
among deuterons and excess neutrons becomes much larger than the 
deuteron radius, the excess neutrons can not excert any significant
influence on the deuteron wave function. 
This is quite opposite to the weak-coupling case $\mu \gg \de$ 
considered above, 
where we have seen that only a slight neutron excess destroys the 
$np$ Cooper pairs. 

For negative chemical potentials Eq.~(\ref{e:wcaa}) is not valid,
since the Fermi surface drops into the unphysical region. 
However, the low-density limit $\rho \ra 0$ of the BCS equations 
provides simple analytical expressions, 
once the density is so low that the chemical potential is close to the
asymptotic value $\mu \ra -\mu_D = -E_D/2 \approx -1.1$ MeV.
%where $E_D$ is the deuteron binding energy.
In some sense now the gap equation (\ref{e:gap0}) and the density 
equations (\ref{e:rho0},\ref{e:drho0}) interchange their roles, 
because the gap equation goes over into the Schr\"odinger equation, 
determining the chemical potential, whereas the value of the gap can be 
extracted from the equations for the density, in the following manner:
%In this situation 
The density of protons is given by
\bea
  \rho_p &=& { (2m)^{3/2} \over 2\pi^2 }
  \int_{e_{\rm min} \ll |\mu|}^\infty de \sqrt{e} \;
  {1\over2} \Bigg[ 1 -  {e-\mu \over \sqrt{(e-\mu)^2 + \de^2} } \Bigg] 
\nonumber\\
  &\approx& { (2m \mu_D)^{3/2} \over 2\pi^2 }
  {\de^2\over 4 \mu_D^2}
  {\int_0^\infty dx {\sqrt{x} \over (x+1)^2 }}  \:.
\eea
The numerical value of the last integral being $\pi/2$, one obtains for the
asymmetry
\be
 \alpha = { \rho-2\rho_p \over \rho } =
 1 - {\rho_\mu \over \rho}{\de^2 \over \mu_D^2}  
\ee
with
\be 
 %\quad ,\quad
 \rho_\mu \equiv { (2m\mu_D)^{3/2} \over 8\pi }
 \approx 0.00049\,\rm fm^{-3}  \:.
\ee
The low-density result is therefore \cite{FAY}
\be 
 {\de\over\mu_D}(\rho,\alpha) = \sqrt{ \rho(1-\alpha) \over \rho_\mu } \:,
\label{e:fay}
\ee
and thus a gap exists for any asymmetry.

%In the limit of maximum asymmetry $\alpha \ra 1$, i.e., $\rho \ra \drho > 0$,
%the gap is vanishing according to Eq.~(\ref{e:fay}) and the system goes over
%into an extremely diluted deuteron gas embedded in a free neutron gas with
%density $\delta\rho$. 
%The Bose-condensed deuterons occupy the negative energy states and the free 
%neutrons the positive ones. 
%With increasing asymmetry the neutrons accomodate in the next positive energy 
%states according to the Pauli principle.
%In this limit Eq.~(\ref{e:schr}) becomes quite similar to the deuteron 
%Schr\"odinger equation except for the blocking of the positive energy states 
%occupied by the free neutrons. 

The ineffectiveness of the neutrons in the $\rho \ra 0$ limit also becomes 
evident when considering Eq.~(\ref{e:schr}).
Indeed in the limit $\rho_n,\rho_p \ra 0$ also 
the momentum distribution $n_k$
%$f(E^+_k)$ and $f(E^-_k)$
vanishes and then the equation goes over into the free deuteron  
Schr\"odinger equation, independent of the asymmetry of the system.
Since the Pauli blocking becomes less and less important as the
density decreases, the asymmetry can become larger
without destroying the superfluidity. 
As we mentioned already in the introduction, this aspect can become important 
in the far tail of the nuclear densities or in other low-density nuclear 
systems. 

In the presence of a neutron excess the Bose-condensed deuterons occupy the 
negative energy state $E \approx -E_D$ and the free neutrons the 
positive energy states. 
With increasing asymmetry the neutrons accomodate in the next positive
energy states according to the Pauli principle.
As we see from the wave function in the introduction, even in the low-density 
limit the excess neutrons stay antisymmetrized with the neutrons bound in 
the deuterons. 
Therefore one cannot distinguish between bound and unbound neutrons, 
the chemical potential of the neutrons is always the one of the unbound ones 
which is tending to zero as $\rho$ approaches zero. 
On the other hand the proton chemical potential $\mu_p$ tends to the binding 
energy of the deuteron, 
since it is the binding energy of the system per half the number of 
particles bound in the deuterons.
Therefore the mean chemical potential $\mu=(\mu_n+\mu_p)/2$ tends to half the 
binding energy of the deuteron such that the eigenvalue $2\mu$ of 
Eq.~(\ref{e:schr}) hits precisely the eigenvalue of the deuteron at $\rho=0$.

To summarize, the behavior of the various chemical potentials in the 
low-density limit is
$\mu_n \ra 0$, $\mu_p \ra -E_D$, $\mu \ra -E_D/2$, $\dmu \ra E_D/2$,
independent of the asymmetry!

%------------------------------------------------------------------------------
\section{Numerical Results}
\label{s:num}

We discuss now the results that were obtained by numerical solution
of the system of equations (\ref{e:eq0}), 
using the Argonne $V_{14}$ potential as the bare interaction.

\begin{figure}
\includegraphics[totalheight=12.cm,angle=270,bb=150 -80 550 480]{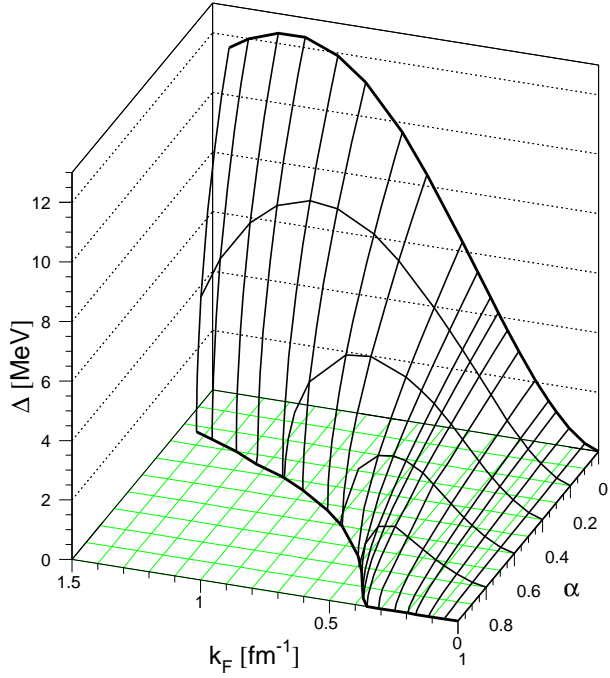}
\caption{
The pairing gap as a function of total density 
[$k_F \equiv (3\pi^2\rho/2)^{1/3}$]  
and asymmetry.}
\label{f:gap}
\end{figure}

We begin, in Fig.~\ref{f:gap}, with an overview of the resulting gap in the
density-asymmetry plane.
As discussed in section~\ref{s:zero},  in the high-density
pairing regime the superfluidity vanishes at a 
finite asymmetry $\al_{\rm max}(\rho)$,
decreasing with increasing density,
whereas in the low-density bound-state region 
(below $k_F \approx 0.35\;\rm fm^{-1}$, $\rho \approx \rho_0/100$)
a gap exists for any asymmetry,
roughly following the analytical result, Eq.~(\ref{e:fay}).

\begin{figure}
\includegraphics[totalheight=12.cm,angle=270,bb=50 70 550 630]{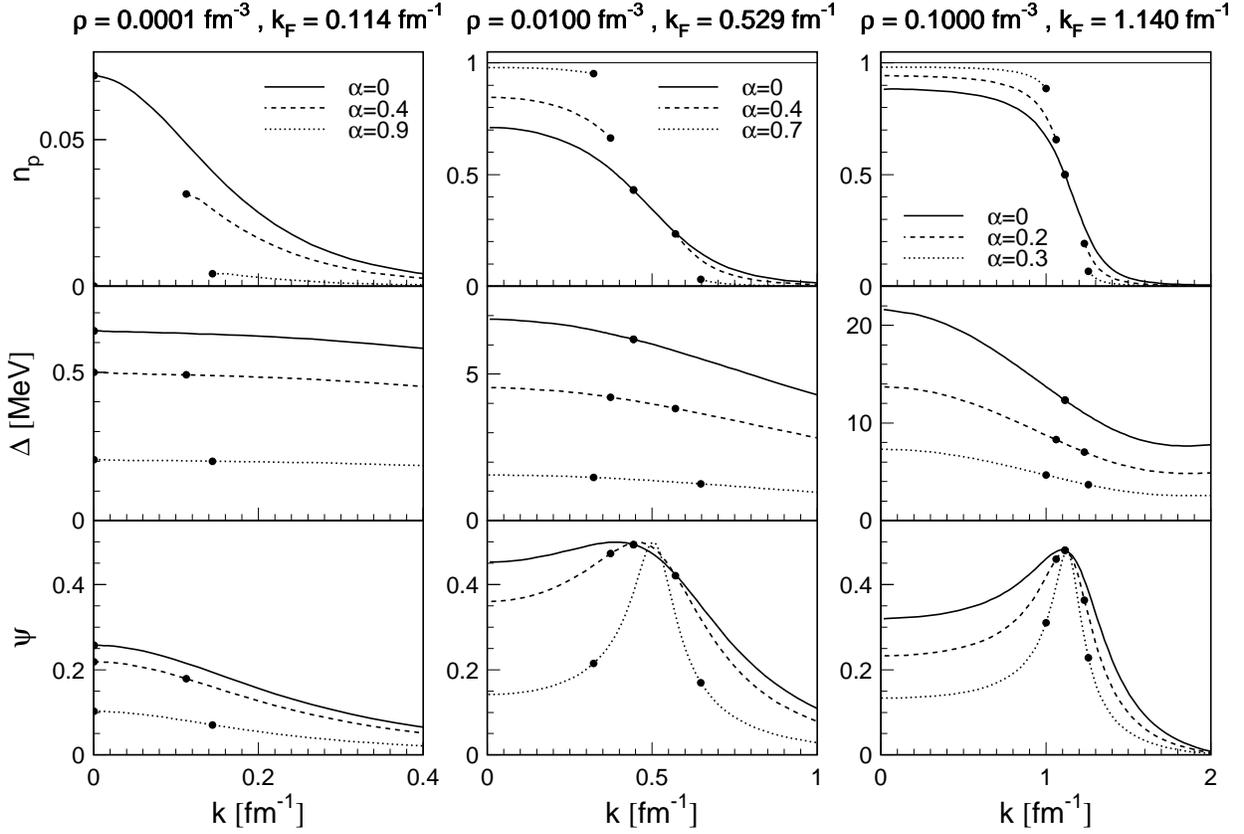}
\caption{
Proton momentum distributions $n_p(k)$, gap functions $\de(k)$,
and anomalous densities $\psi(k)$,
for different total densities $\rho=10^{-4}, 10^{-2}, 10^{-1}\;\rm fm^{-3}$, 
and different asymmetries.
The dots indicate the boundaries of the interval containing the neutron
excess in each case.}
\label{f:n}
\end{figure}

In Fig.~\ref{f:n}, we show the momentum distributions,
gap functions, and anomalous densities at three different densities
$\rho=10^{-4}, 10^{-2}, 10^{-1}\;\rm fm^{-3}$, 
representative of the low, intermediate, and high-density 
regions mentioned in section~\ref{s:zero}.
The top panels of the figure show the momentum distributions
of the superfluid protons.
As discussed before, see Fig.~\ref{f:dis},
there is a proton-free region that is at high density
centered around the Fermi momentum, and at low density excludes all
low-momentum states, representing the Fermi sea of non-superfluid excess 
neutrons.
The resulting variation of the gap function in the relevant regions of 
momentum space is rather small at low density, however, it is competing
with the kinetic energy in the self-consistent determination of
the normal region, which is determined by the variation of 
$E_k=\sqrt{\eps_k^2+\de_k^2}$.
The anomalous density develops a characteristic peak at the 
Fermi momentum only in the high-density regime, 
whereas at low density it exhibits a smooth variation typical of the
deuteron wave function.

\begin{figure}
\includegraphics[totalheight=13.cm,angle=0,bb=-80 30 -80 720]{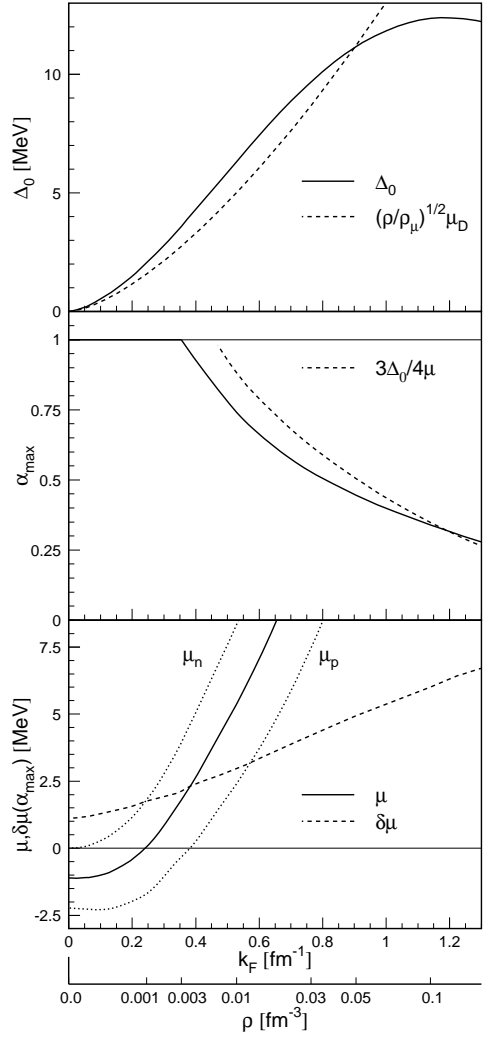}
\caption{
Top panel: The $^3SD_1$ gap in symmetric nuclear matter as a function of total
density $\rho$ or equivalent Fermi momentum $k_F = (3\pi^2\rho/2)^{1/3}$ 
(solid line). 
The dashed line shows the analytical approximation Eq.~(\ref{e:fay}).
Central panel: The maximum asymmetry at which a gap exists
(solid line) and the approximation Eq.~(\ref{e:amax}) (dashed line).  
Lower panel: The values of the various chemical potentials 
$\mu,\dmu,\mu_n,\mu_p$, corresponding to the asymmetry $\al_{\rm max}$ 
displayed in the panel above.}
\label{f:mu}
\end{figure}

In Fig.~\ref{f:mu}, we show in more detail the gap in symmetric matter,
together with the analytical approximation Eq.~(\ref{e:fay}) (top panel);
the maximum value of asymmetry at which the superfluidity disappears,
in comparison with the analytical estimate Eq.~(\ref{e:amax}) (middle panel);
as well as, in the bottom panel, 
the variation of the various chemical potentials along the line
$[\rho,\al_{\rm max}(\rho)]$.
The asymptotic behavior predicted in the previous section,
$(\dmu,\mu_n,\mu,\mu_p) \ra (1,0,-1,-2) \mu_D$,
is observed for $\rho\ra0$.  

\begin{figure}
\includegraphics[totalheight=7.cm,angle=270,bb=160 -180 560 380]{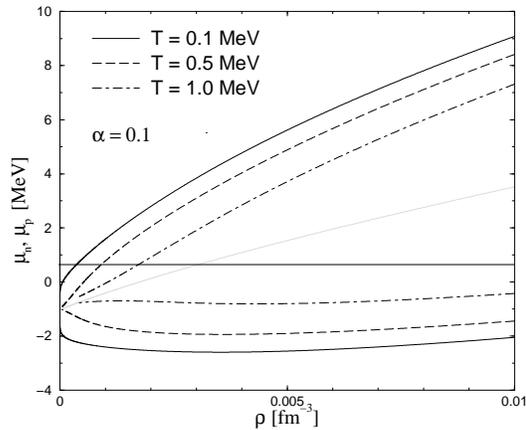}
\caption{
Chemical potentials of neutrons (upper curves) and protons (lower curves)
at $\al=0.1$ as functions of density for different temperatures.}
\label{f:t}
\end{figure}

We finally discuss briefly the situation at finite temperature, that
was treated in detail in Refs.~\cite{ALM,SCK,ARM1,ARM2}.
As long as $T\ll\de$, the temperature effects are small.
However, in the low-density limit, when $\de \ra 0$, a qualitatively 
different behavior from the zero temperature case can be observed:
In Fig.~\ref{f:mu} we have seen that in this limit
$\mu_n \ra 0$ and $\mu_p \ra -2\mu_D$,
due to the %presence of 
coexistence of the deuterons with 
a Fermi sea of free neutrons,
see the discussion in section~\ref{s:zero}(i).
At finite temperature, however, part of the deuterons are broken up and 
therefore this argument does not apply any more.
%there is no reason then that $\mu_n$ and $\mu_p$ should be different.
This can be seen from Eq.~(\ref{e:drho}) for $\drho$.
In the limit $\rho,\drho \ra 0$ we have $\de = 0$,
since at finite temperature the system becomes normal at a certain
critical density.
Therefore Eq.~(\ref{e:drho}) reads
\be
 0 = 2\sum_k \left[ f(\eps_k-\dmu) - f(\eps_k+\dmu) \right] \: 
\ee
and one clearly sees that this relation can only be fulfilled,
at any finite temperature, for $\dmu=0$, i.e., $\mu_n=\mu_p$.
On the other hand, we see from Eq.~(\ref{e:schr}) that $\mu \ra -\mu_D$.
Therefore we have for $T \ne 0$ the limit $\mu,\mu_n,\mu_p \ra -\mu_D$.
This behavior is clearly born out from the numerical calculation,
Fig.~\ref{f:t},
which shows the density dependence of $\mu_n$ and $\mu_p$
for fixed asymmetry $\alpha=0.1$ and various temperatures 
$T=0.1,0.5,1.0\;\rm MeV$.
It is seen how with vanishing temperature the situation shown in the 
lower panel of Fig.~\ref{f:mu} (for $\al=1$) is approached.

%------------------------------------------------------------------------------
\section{Conclusions}
\label{s:conc}

%In this article we have investigated cold asymmetric nuclear matter within
%the BCS theory.
%It was shown that in the high-density pairing regime the superfluid phase
%is restricted to a narrow asymmetry interval, 
%whereas in the dilute system a finite gap persists for all asymmetries.
%In this situation the gap equation describes a system of Bose-condensed
%deuterons coexisting with a Fermi gas of free neutrons, 
%which is Pauli-blocking the deuteron bound states.
%Analytical approximations for the gap in the low and high density regimes were
%given, and the resulting behavior of the chemical potentials was analyzed.

In this article we extended a previous study \cite{SCK} of the transition
from a neutron-proton BCS superconducting state to a Bose condensate of
deuterons in symmetric nuclear matter to systems with isospin asymmetry.
This is an important aspect, since most of the low-density nuclear systems,
such as tails of nuclear density distributions in nuclei, have a strong neutron
excess.
In the high-density weak-coupling regime, $\mu \gg \de$,
even a small asymmetry is sufficient to suppress 
the pairing correlations completely via the Pauli blocking effect. 
However, we find that in the situation where the density drops so low that 
the chemical potential turns negative and deuterons start to form, 
the isospin asymmetry is much less important, i.e., the system supports 
pair correlations (in the form of a deuteron condensate) 
for much larger asymmetries.

Indeed one can argue that, once the density is so low that the spatial 
separation between deuterons and between deuterons and extra neutrons is large,
the Pauli principle is ineffective and, hence, asymmetry can not destroy any 
longer the binding of neutron-proton pairs.  
Our numerical calculations fully confirm this conjecture. 
Most easily this effect can be understood by examining the phase space 
blocking effect on the anomalous density.
In the high-density regime the latter quantity shows a peak structure as a 
function of the mean chemical potential of neutrons and protons $\mu$. 
Once the asymmetry is imposed the Pauli blocking effectively cuts out the 
major part of this peak structure.
When the chemical potential turns negative, there is no peak any more
in the relevant physical region of the anomalous density and the Pauli blocking
looses its efficiency, which enables the proton-neutron pairs to condense 
in the very low density regime even in the presence of a large neutron excess.
We anticipate that this aspect may be important for the understanding of
the far tails of density profiles of exotic nuclei, the expanding (asymmetric)
nuclear matter in heavy ion collisions, and other low-density nuclear systems.

\section*{Acknowledgements}
This work has been partially supported by the 
``Groupement de Recherche: Noyaux Exotiques, CNRS-IN2P3, SPM,"
as well as by the programs
``Estancias de cient\'{\i}ficos y tecn\'ologos extranjeros en Espa\~na,''
SGR98-11 (Generalitat de Catalunya), and DGICYT (Spain) No.~PB98-1247.

%------------------------------------------------------------------------------

%------------------------------------------------------------------------------

\end{document}